\documentclass[11pt,a4paper]{article}
\usepackage[latin9]{inputenc}
\usepackage{amsmath}
\usepackage{amssymb,color}
\usepackage{graphicx}
\usepackage{esint}
\usepackage{cite}
\makeatletter
\usepackage{subfigure}
\usepackage{amscd}
\usepackage{appendix}
\usepackage{verbatim}
\usepackage{amsmath,bm}

\begin{document}
	\begin{titlepage}		{\hbox to\hsize{\hfill{} CTPU-PTC-22-21 }}
		
		\bigskip{}
		\vspace{3\baselineskip}
		
		\begin{center} 
			\textbf{\Large{Resonant Vector Dark Matter Production during Inflation}}\par
		\end{center}{\Large \par}

		\vspace{0.1cm}
		\begin{center}
			\textbf{
				Neil D. Barrie
			}\\
			\textbf{ }
			\par\end{center}
		
		\begin{center}
			{\it Center for Theoretical Physics of the Universe, Institute for Basic Science (IBS),\\ 
			 Daejeon, 34126, Korea\\ 
	Email: nlbarrie@gmail.com}\\
			\textit{\small{}}
			\par\end{center}{\small \par}
		
		\begin{center}
			\textbf{\large{}Abstract}
			\par\end{center}{\large \par}
		
		\noindent 
We present a new mechanism for generating the observed dark matter energy density, through resonant production of vector dark matter during inflation. Resonant amplification of the vector dark matter is achieved through Mathieu instabilities induced by an oscillatory coupling with the inflaton, which can occur as early as 11.9 e-folds before the end of inflation. Successful dark matter production is possible for a wide range of inflationary Hubble rates and vector mass scales. This scenario generates both helicity states of the vector dark matter density in contrast to other tachyonic production mechanisms, while exhibiting large parameter regions with negligible energy density generated in the longitudinal mode. The discovery of this dark matter candidate will not only explain the nature of dark matter but provide insight into the inflationary potential and the scale of inflation. The allowed parameter space is surveyed and the unique phenomenological implications are discussed, which include gravitational waves and the dark matter energy density spectrum.
		
\end{titlepage}
	
\section{Introduction}

A major mystery of modern physics is the nature and origin of dark matter. From observation, it is known that the dark matter energy density of the universe today is approximately \cite{Aghanim:2018eyx},
\begin{equation}
\rho_{\textrm{DM}} = 9.6\cdot 10^{-48}~\textrm{GeV}^4 ~.
\label{DM_today_obs}
\end{equation}
Despite the increased sensitivity and breadth of experiments searching for the constituents of dark matter, there is yet to be convincing experimental evidence for the existence of any proposed dark matter candidates so far. This has led to consideration of new dark matter candidates beyond the usual Weakly Interacting Massive Particle (WIMP) paradigm \cite{Bertone:2016nfn,Roszkowski:2017nbc} - in particular, vector dark matter \cite{Hambye:2008bq,Nelson:2011sf,Arias:2012az,Graham:2015rva,Agrawal:2018vin,Dror:2018pdh,Co:2018lka,Bastero-Gil:2018uel,Long:2019lwl,Ema:2019yrd,Nakayama:2020rka,Nakai:2020cfw,Ahmed:2020fhc,Kolb:2020fwh,Nomura:2020zlm,Gross:2020zam,Co:2021rhi}. An important opportunity presented by alternative dark matter candidates, such as vector dark matter, is that they allow for novel production methods that provide additional observation signatures, leading to extra avenues for testing and confirmation. One particularly intriguing implication of vector dark matter is the possibility of kinetic mixing between this dark photon and the usual photon, observation of which is actively being pursued \cite{Fabbrichesi:2020wbt,Caputo:2021eaa}.

In recent years, the properties and phenomenology of vector dark matter have been an active area of investigation. One interesting prospect is the possible connection between the generation mechanism of the vector dark matter and the inflationary scenario, which provides unique observational phenomenology and may allow probes of the very early universe dynamics. Interestingly, in high-scale inflationary scenarios 	\cite{Guth:1980zm,Linde:1981mu,Albrecht:1982wi,Mukhanov:1981xt,Martin:2013tda} the longitudinal component of a vector field, with a Stueckelberg mass term \cite{Green:1984sg}, is excited by quantum fluctuations that can be sufficiently large to explain the observed dark matter energy density \cite{Graham:2015rva}. Another possibility for generating the vector dark matter is through the $ \varphi F\tilde{F} $ scalar coupling to vector dark matter fields, which may exhibit additional experimental signatures from an associated axion. This coupling can induce tachyonic exponential amplification of one helicity state \cite{Bastero-Gil:2018uel}, as considered in Baryogenesis mechanisms \cite{Alexander:2011hz,Barrie:2014waa,Barrie:2015axa,Anber:2015yca,Papageorgiou:2017yup,Jimenez:2017cdr,Domcke:2019mnd,Barrie:2020kpt,Barrie:2021orn}, as well as generate dark matter through the misalignment mechanism \cite{Nelson:2011sf,Arias:2012az}.

The use of resonant enhancement of electromagnetic fields during inflation has been proposed to achieve successful Magnetogenesis \cite{Byrnes:2011aa}. This scenario involved an oscillatory coupling between the inflaton and the gauge fields, leading to the development of Mathieu instabilities.  Unfortunately, this model is unable to successfully explain Magnetogenesis because the required energy density of the electromagnetic fields generated early in the inflationary epoch is too large and would disrupt the background evolution. However, if one does not require features on large length scales today, as in Magnetogenesis, we can instead generate significant energy densities through these resonances at times closer to the end of inflation without jeopardising the inflationary setting. With this in mind, we propose that this Magnetogenesis mechanism could instead explain the observed dark matter energy density, by coupling the inflaton to a massive vector dark matter particle instead of the electromagnetic field. Providing a new way to connect the origin of dark matter and inflation, while exhibiting unique phenomenology in comparison to previously considered inflationary vector dark matter scenarios.

In this work, we explore the possibility that the observed dark matter density is composed of a massive vector field that was produced resonantly during inflation. The paper will be structured as follows; Section \ref{InfDyn} will describe the dynamics of the vector field and its coupling to the inflaton during inflation, with details of the resonance behaviour and generated energy density. In Section \ref{Gen_VDM}, the evolution of the generated dark matter energy density will be discussed and the regions of allowed parameter space for successful production of the observed relic density presented. Section \ref{phenom} investigates the unique phenomenological signatures of this mechanism compared to other vector dark matter models. Finally, in Section \ref{conc}, we will conclude with a discussion of the implications of these results and future directions for investigation.


\section{Dynamics of the Vector Dark Matter and the Resonance Behaviour during Inflation}
\label{InfDyn}

To begin, we write the action that describes our vector dark matter, or dark photon, candidate $ X_\mu $~,
\begin{equation}
S= \int \textrm{d}^4x \sqrt{-g} \left(\frac{1}{4} X_{\mu\nu}X^{\mu\nu} +m^2 X_\mu X^\mu\right)+S_\textrm{int}~,
\label{action_tot}
\end{equation}
where  $X_{\mu\nu}$  denotes the gauge field strength tensor of $ X_\mu $~, with corresponding coupling constant $ g_{D} $~, $ S_\textrm{int} $ describes its interaction terms, and $ m $ is the mass parameter of the vector dark matter, which may be a Stueckelberg mass term. It is not necessary for this mass term to be of the Stueckelberg type, or to be present during inflation, and can instead be generated after reheating through spontaneous symmetry breaking, as long as this occurs sufficiently before matter-radiation equality. For the analysis that follows, we require at the very least that $ m\ll H_{\textrm{inf}} $~, where  $ H_{\textrm{inf}} $ is the approximately constant Hubble rate during inflation, so that it does not affect the resonance behaviour. In all cases considered below, it is assumed that the vector field mass plays no role during inflation.  

The interaction term we consider between the vector dark matter field and the inflaton $ \varphi $ has the following form,
\begin{equation}
S_\textrm{int} = \int \textrm{d}^4x ~\sqrt{-g}~ \frac{1}{4}f(\varphi) X_{\mu\nu}\tilde{X}^{\mu\nu} ~ ,
\label{action_int}
\end{equation}
where the dual of the field strength tensor is defined as $\tilde X^{\mu\nu}= \frac{1}{2\sqrt{-g}}\epsilon^{\mu\nu\rho\sigma} X_{\rho\sigma}$~. The early universe implications of this type of coupling have been widely studied, in particular concerning mechanisms for Baryogenesis \cite{Alexander:2011hz,Barrie:2014waa,Barrie:2015axa,Anber:2015yca,Jimenez:2017cdr,Domcke:2019mnd,Barrie:2020kpt}, Gravitational Leptogenesis \cite{Papageorgiou:2017yup,Barrie:2021orn}, and Dark Matter production \cite{Barrie:2015axa,Bastero-Gil:2018uel}, in which  the coupling is typically taken to be of the form $ f(\varphi)=\frac{\varphi}{\Lambda} $ with $ \Lambda $ being a UV cutoff scale. Interestingly, the Gravitational Leptogenesis mechanisms involving gauge fields can also produce vector dark matter remnants, a connection to be explored in our future work. 
Our mechanism does not require a kinetic mixing to the usual photon, and if it exists, it is assumed that it is small enough that it can be ignored during the inflationary setting.

As mentioned in the Introduction, the longitudinal component can contribute significantly to the dark matter energy density if the inflationary setting is sufficiently high-scale. In our scenario, the coupling introduced in Eq. (\ref{action_int}) does not alter the dynamics of the longitudinal component, so its dynamics and resulting energy density will be identical to the analysis in Ref. \cite{Graham:2015rva}. Thus, we analyse the dynamics of the transverse components, which have the following equation of motion for the action given in Eq. (\ref{action_tot}) and (\ref{action_int}), 
\begin{equation}
\left(\partial_{\tau}^2-\vec \bigtriangledown^2 +a(\tau)^2 m^2 \right) X^{i}+\frac{d f(\varphi)}{d\tau}\epsilon^{ijk}\partial_j X_k=0~,
\label{Ch3xfieldeom4}
\end{equation}
where we have taken the coulomb gauge  $(X_\mu)=(0,X_i)$ with $\partial_iX_i=0$. To allow analytical treatment, we will make the simplifying assumption that the back-reaction on the motion of $ \varphi $ due to the production of the vector field $ X_i $ is negligible, and that the energy density in the vector field is never greater than $ 5\% $ of the total energy density during inflation.

To  quantize this model, we promote the $ X $ vector fields to operators and assume that the vector dark matter has two possible circular polarisation states,
\begin{equation}
X_i=\int\frac{d^3\vec k}{(2\pi)^{3/2}}\sum_{\alpha}\left[F_{\alpha}(\tau,k)\epsilon_{i\alpha}\hat a^a_{\alpha} {\rm e}^{i\vec k\cdot\vec x}+
	F^{*}_{\alpha}(\tau,k)\epsilon^{*}_{i\alpha}\hat a_{\alpha}^{a\dagger}{\rm e}^{-i\vec k\cdot\vec x}
	\right]~,
	\label{Ch3112}
\end{equation}
where $\vec \epsilon_{\pm}$ denotes the two possible helicity states of the $X$ vector field ($\vec \epsilon_{+}^{*}=\vec \epsilon_{-}$), while the creation, $\hat a_{\alpha}^{\dagger}(\vec k)$, and annihilation, $\hat a_{\alpha}(\vec k)$, operators satisfy the canonical commutation relations,
\begin{equation}
	\left[\hat a_{\alpha}(\vec k), \hat a_{\beta}^{\dagger}(\vec k')\right]=\delta_{\alpha\beta}\delta^3(\vec k-\vec k')~,
\end{equation}
and      
\begin{equation}
	\hat a^a_{\alpha}(\vec k)\vert 0\rangle_{\tau}=0~,
	\label{Ch3145}
\end{equation}
where $\vert 0\rangle_{\tau}$ is an instantaneous vacuum state at time $\tau$.

Finally, substituting into the equations of motion in Eq. (\ref{Ch3xfieldeom4}), we find that the mode functions in Eq. (\ref{Ch3112}) are described by the following equation,
\begin{equation}
	F^{''}_\lambda + \left(k^2  +a(\tau)^2 m^2 \pm k \frac{d f(\varphi)}{d\tau}\right) F_\lambda = 0~,
	\label{mode_func_vdm}
\end{equation}
where $ \lambda=+,- $ denotes the helicity states of the vector field, and primes denote derivatives with respect to $ \tau $.

Keeping the form of the above equation for the wave mode functions in mind, consider the well-known Mathieu equation,
\begin{equation}
\frac{\textrm{d}^2 y(z)}{\textrm{d} z^2} + \left( p - 2q \cos 2z \right) y(z) = 0 ~.
\label{mathieu}
\end{equation}
An intriguing feature of the Mathieu equation is that it permits solutions that exhibit instabilities, that is, exponentially growing solutions. This property has been utilised in the context of axion oscillation induced resonances to produce vector dark matter after reheating \cite{Agrawal:2018vin}. 

Now, if we wish for our equation of motion in Eq. (\ref{mode_func_vdm}) to exhibit such solutions, the parameters and couplings of our model must fit the following forms,
\begin{equation}
p = \frac{k^2 +a(\tau)^2 m^2}{\omega^2}  ~ , ~~\textrm{and} ~~  2 q  \cos 2\omega\tau =\frac{k   }{\omega^2} \frac{d f(\varphi)}{d\tau} ~ ,
\label{relations}
\end{equation}
where it is important to note that we will assume that $ m $ is small compared to modes of interest in $ k $ and $ \omega $ such that it can be neglected in the resonance analysis and throughout inflation.

It is then possible to interpret that, to fit the Mathieu equation, the coupling function of the inflaton to the vector field must be approximately of the form,
\begin{equation}
f(\tau) = \gamma \sin (2\omega\tau) ~ .
\label{coupling}
\end{equation}
The relation between the values of $ p $ and $ q $ will then determine the existence and properties of the instabilities that occur. In Ref. \cite{Byrnes:2011aa}, it is discussed how a coupling of this form could arise in the inflationary setting. One way is to begin with a successful inflationary background model and determine the coupling function $ f(\varphi) $ that results in the required functional form to produce oscillatory behaviour from the slow rolling inflaton field. Namely,
\begin{equation}
f(\varphi) = \int \frac{d f(\tau)}{d\tau} \frac{d\tau}{d\varphi} d\varphi ~ .
\end{equation}
For example, this coupling is easily constructible in the context of the chaotic and hybrid inflation inflationary settings, although their form may not appear particularly natural \cite{Byrnes:2011aa}.

On the other hand, it is possible to proceed in the opposite direction, starting with a simple form for the coupling  $f(\varphi)$ from which the required inflaton dynamics are determined, while requiring successful inflation. Interestingly, periodic oscillations in the inflaton evolution can exist in inflationary scenarios such as Axion Monodromy models \cite{Pahud:2008ae}. Although, any oscillations must be sub-dominant to ensure the stability of the background dynamics, which may make it difficult to produce the required resonance behaviour. A coupling of the form $f(\Box \varphi)$ is likely to be more successful in this case, as $\ddot{\varphi}$ is allowed to oscillate \cite{Byrnes:2011aa}.

If we assume that the coupling takes the form given in Eq. (\ref{coupling}), we can analyse the resonant behaviour and its relation to the different parameters of the model. To illustrate this, we show the solution for the constant frequency $ \omega $ case. In solving this equation, for certain relations between $ p $ and $ q $ there exist exponentially growing solutions. This occurs for modes $ k=n\omega $ with $ n\in\mathbb{N} $, with solutions of the following approximate form,
\begin{equation}
|F_+(\tau)|^2 + |F_-(\tau)|^2 \simeq |F_\pm(\tau_i)|^2  e^{2\mu\omega(\tau-\tau_i)},
\label{Const_freq_case}
\end{equation}
for $ k \simeq n\omega $ with $ n $ an integer, and where $\tau-\tau_i $ is the length of time over which the resonance occurs and we assume the same initial conditions for both helicities. Note that due to the opposite sign of the oscillatory coupling for each of the two helicities, they are out of phase by $ \pi/2 $. The $ \mu $ parameter is  known as the Floquet index, which is k-dependent, but can be well approximated by $ \mu\simeq \gamma/2 $. The resonance band around these modes is determined by the relation between $ p $ and $ q $, where in our case $ q=\gamma k/\omega=\gamma \sqrt{p} $, and thus $ \gamma $ is the important parameter for the width in our context. 

If this behaviour exists for a fixed mode for the entirety of inflation, the exponential enhancement would be dependent on initial conditions and likely disrupt the inflationary background. If instead the frequency $ \omega $ is time-dependent, different modes will be excited during inflation at varying times. Here we follow the analysis in Ref. \cite{Byrnes:2011aa} as a proof of concept for our model, in which the frequency in the coupling function is taken to be,
\begin{equation}
\omega(\tau)=k_* e^{k_0 \tau} ,
\label{time_varying}
\end{equation}
such that the time-varying frequency behaviour is imposed, with $ k_0 $ and $ k_* $ characteristic modes satisfying $ k_*\gg k_0 $~. Importantly, this choice also alleviates the issue of initial conditions associated with the constant frequency case, allowing the standard Bunch-Davies initial state for the wave mode functions. Additionally, there is a finite period in which a mode enters into a resonance band and experiences exponential enhancement. The resonance band entrance and exit times are given by,
\begin{equation}
\tau_\pm=\frac{1}{k_0}\ln\left[ \frac{k}{(1\pm\gamma/2) k_*}\right]~,
\end{equation}
where this period is approximately given by $ \Delta \tau\simeq \frac{\gamma}{k_0} $ for small $ \gamma $. This shows the need for $ k_0\ll k_* $~, to ensure that there are sufficient oscillations within the resonance period.

The Floquet index varies across the resonance band, with numerical calculations giving the following approximate exponential enhancement factor,
\begin{equation}
|F_+|^2+|F_-|^2\simeq |F_\textrm{initial}|^2 e^{\frac{\pi\alpha}{8}\frac{k}{k_*} }~,
\label{resonance}
\end{equation}
where we define the parameter $ \alpha=\gamma^2\frac{k_*}{k_0} $ ~, and $ F_\textrm{initial} $ is taken to be the Bunch-Davies vacuum state. Only modes within the range $ k_0<k<k_* $ experience resonant enhancement. Notice the $ k $ dependence in the exponent, which leads to a strongly blue shifted energy density spectrum, dominated by the characteristic mode $ k_* $~. 

From this, we can calculate the approximate dark matter energy density generated through the resonance by horizon crossing for the last mode that experiences resonance enhancement, $ k_* $ . This occurs at $ \tau=\tau_* $ , and corresponds to the following dark matter energy density, 
\begin{equation}
\rho_{\textrm{DM}}^*=\frac{1}{2 a^4(\tau_*)}\int^{k_*}_{k_0} \frac{d \textbf{k}^3}{(2\pi)^3} k^2 (|F_+|^2+|F_-|^2)~,
\label{rhoVDM}
\end{equation}
where this is the corresponding dark magnetic field energy density, and we define $ k_*|\tau_*|=1 $.

We focus on the characteristic frequency $ k_* $ , which will give the largest contribution to this energy density due to the $ k $ dependence of the exponential growth factor. If this scale is of the order of the inflationary Hubble rate, then $\tau_*$ will coincide with the end of inflation, $\tau_\textrm{end}$~, where we define $ a(\tau_\textrm{end})=1 $. If instead, we take $ k_* <H_{\textrm{inf}} $~, then the resonance dynamics will instead end at the horizon crossing for the mode $  k=k_* $~, i.e. $ k_*|\tau_*|=1 $. Thus, we must also take into account the subsequent inflationary dilution of the vector dark matter energy density after the excited modes exit the horizon. We define $ N_* $ as the number of e-folds before the end of inflation that the mode $ k_* $ exits the horizon. 

In the case of small $ \gamma $, and taking into account variation of the Floquet index across the resonant band, the resultant energy density in the vector field at the end of inflation is then given by,
\begin{equation}
\rho_{\textrm{DM}}^\textrm{end} \simeq \frac{ k_0k_*^3}{\pi^3 \gamma^2} 
e^{\frac{\pi\gamma^2k_*}{8k_0}}  =\frac{ k_*^4}{\pi^3 \alpha} 
e^{\frac{\pi\alpha}{8}}~,
\label{rhoVDM1}
\end{equation}
where the region of validity of this solution is $ 1\ll \frac{k_*}{k_0}\lesssim 10^4 $ \cite{Byrnes:2011aa}.

It is important to check that all the $ k $ modes that are resonantly enhanced never dominate the energy density of the inflaton, which is ensured if true for the $ k=k_* $ mode. We find this to be easy to satisfy for successful replication of the observed dark matter energy density in the $ k_*=H_\textrm{inf} $ scenario. However, in considering the $ k_*<H_\textrm{inf} $ case, it is important to determine when the resonance ends for the mode $ k_* $~. If it underwent exponential growth several e-folds before the end of inflation, we must check that the resultant energy density produced did not compete with that of the inflaton and thus destroy the inflationary dynamics. That is, we require the following,
\begin{equation}
\rho_{\textrm{DM}}^* \simeq \frac{H_{\textrm{inf}}^4}{\pi^3 \alpha}
e^{\frac{\pi\alpha}{8}}  \ll 3M_p^2 H_{\textrm{inf}}^2  ~,
\label{rhoVDM*}
\end{equation}
where the $ * $ represents evaluation at the end of the resonance. From this, we can derive an upper bound on the allowed $ \alpha $ parameter, $ \alpha \lesssim 198 $, when requiring $ H_{\textrm{inf}}>10^3 $ GeV, assuming the vector dark matter makes up less than $ 5\% $ of the total energy density.  The earliest time in inflation that this resonant enhancement can safely occur will be discussed in the next Section. 
 
Importantly, although we have used the above oscillatory coupling to induce the instabilities, it is not necessary that this be the exact form of the coupling or that an exact Mathieu resonance appears. We have used this example to demonstrate the ease with which this mechanism can generate the observed dark matter energy density, and thus, less efficient oscillatory coupling scenarios should also be able to succeed.


\section{Generated Vector Dark Matter and Evolution after Inflation}
\label{Gen_VDM}

Now that we have determined the dynamics of the resonance amplification that occurs during inflation and the resultant dark matter energy density generated, we proceed to the calculation of the corresponding relic abundance of the vector dark matter today. Thus, finding the allowed parameter regions that are consistent with observation, as given in Eq. (\ref{DM_today_obs}). We have assumed the standard cosmological evolution after reheating, but if there were periods of kination or early matter dominated phases the allowed coupling ranges may be altered for certain parameter ranges. Additionally, we have made the simplifying assumptions that the Hubble rate is constant throughout inflation, and the reheating is instantaneous. Altering these assumptions would change the longitudinal upper bound, and slightly vary the required alpha parameter for a given vector dark matter mass and inflationary Hubble rate.

Due to the momenta dependence of the exponential resonant enhancement, the momentum distribution is strongly peaked at the last mode to be resonantly generated. Thus, once inflation ends, the generated vector dark matter will have a characteristic physical momenta of $ k_* $ . Once reheating occurs, and the universe enters a radiation dominated evolution, this characteristic momenta will redshift as $k_{\textrm{phys}}= k_*\frac{T}{T_{\textrm{RH}}} $ with the total energy density of the vector dark matter diluting as $ a^{-4} $. This behaviour continues until the physical momenta becomes comparable to its mass $k_{\textrm{phys}}\sim m$, after which the energy density evolves as matter-like to today. The transition to a matter-like evolution must occur before matter-radiation equality for the vector dark matter to be a successful dark matter candidate. Following this evolution from the beginning of inflation to today, we are able to determine the required dark matter energy density at the end of inflation to be consistent with observation. 

As noted in the Introduction, it has been found that the inflationary perturbations can excite the longitudinal mode of a massive vector field that has a Stueckelberg mass term. The resulting energy density can be consistent with the observed dark matter energy density in high-scale inflationary scenarios. As this longitudinal mode may also be excited in our model, we will focus on regions of parameter space for which the longitudinal mode would not be sufficiently produced to be the sole component of dark matter. 
To avoid the energy density of the longitudinal modes induced by quantum fluctuations dominating the dark matter, we require that \cite{Graham:2015rva},
\begin{equation}
\frac{\Omega_\textrm{L}}{\Omega_{\rm CDM}} = \sqrt{\frac{m}{6\cdot 10^{-15}~\text{GeV}}} \left(\frac{H_{\textrm{inf}}}{10^{14}~\text{GeV}} \right)^2 \ll 1~ ,
\label{long_con}
\end{equation}
which will apply a constraint on the parameter space for our mechanism, where we assume it always constitutes less than $ 5\% $ of the dark matter energy density today. Thus, our first constraint on the allowed parameter space is,
\begin{equation}
m < 2.6 \cdot 10^{-12}~\text{GeV} \left(\frac{4.9\cdot 10^{12}~\textrm{GeV}}{H_{\textrm{inf}}} \right)^4~.
\label{long_con1}
\end{equation}
It should be noted that this constraint applies if the vector dark matter has a Stueckelberg mass term. In what follows we will include this constraint, but this constraint will disappear if this is not the case. This would allow larger inflationary Hubble rates to be considered.


\subsection{Resonant Production at the End of Inflation: $ k_*=H_\textrm{inf} $}

To begin the analysis, let us consider the $ k_*=H_\textrm{inf} $ case, for which the dark matter energy density generated by the end of inflation is given by,
\begin{equation}
\rho_{\textrm{DM}}^\textrm{end} \simeq \frac{ H_{\textrm{inf}}^4}{\pi^3 \alpha} 
e^{\frac{\pi\alpha}{8}}~.
\label{rhoVDM2}
\end{equation}
For simplicity, we first explore the predictions in this case, before determining the earliest point in inflation, or largest $ N_* $ , for which this mechanism successfully explains dark matter without destabilising the inflationary setting. 

The typical physical momenta of the vector dark matter at the end of reheating will be $ k_{\textrm{phys}}= H_\textrm{inf} $~, assuming instantaneous reheating. The vector dark matter energy density will subsequently dilute as radiation until this momenta becomes comparable to its Compton frequency $ m $. Once this has occurred, it will become non-relativistic and the dark matter energy density will instead redshift in a matter-like way.  The temperature at which this occurs is given by, 
\begin{equation}
T_m =m \frac{T_\textrm{reh}}{H_\textrm{inf}}= m\left( \frac{90}{\pi^2 g_*} \right)^{1/4}  \left(\frac{M_p}{H_\textrm{inf}}\right)^{1/2}~ ,
\label{temp_mass}
\end{equation}
where we have taken $ k_{\textrm{phys}}=H_\textrm{inf}\frac{T}{T_\textrm{reh}} $ in this case. The dark matter energy density at $ T_m $ is then,
\begin{equation}
\rho_{\textrm{DM}}^m = \rho_{\textrm{DM}}^{\textrm{end}} \left( \frac{ T_m}{T_\textrm{reh}} \right)^4 ~,
\label{DMevo1}
\end{equation}
where it is necessary that $ T_m>10^{-9} $ GeV, the approximate temperature at matter-radiation equality, to ensure that our vector field can play the role of dark matter. This requirement places another important constraint on our allowed parameter space,
\begin{equation}
m>2.6 \cdot 10^{-12} ~\textrm{GeV}~  \left(\frac{H_\textrm{inf}}{4.9 \cdot 10^{12}~\textrm{GeV}}\right)^{1/2}~.
\label{temp_mass_con}
\end{equation}

Once the temperature drops below $ T_m $ , the dark matter number density will evolve as non-relativistic matter until today. Thus, the dark matter density today is given by, 
\begin{equation} 
\rho_{\textrm{DM}}^0 = \rho_{\textrm{DM}}^m \left( \frac{T_0}{T_m}  \right)^3~ ,
\label{DMevo2}
\end{equation}
where $T_0 \approx 2\cdot 10^{-13}$ GeV is the current Cosmic Microwave Background (CMB) temperature.

\begin{figure}[t]
	\centering
	\includegraphics[width=0.65\textwidth]{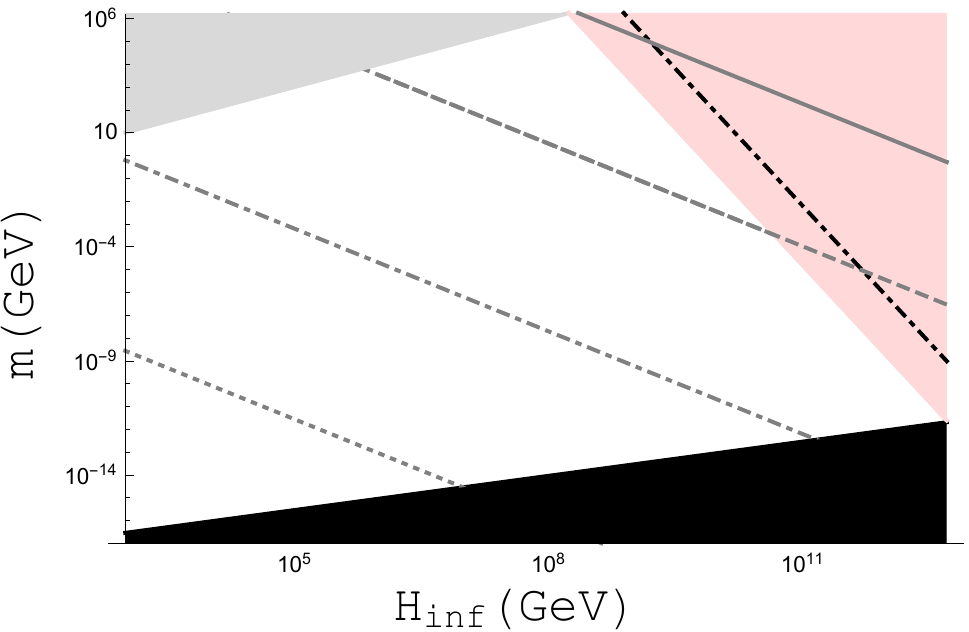}
	\caption{A depiction of the allowed parameter region that generates the observed dark matter energy density today, for the $ k_*=H_\textrm{inf} $ case. The Grey lines indicate different values of $ \alpha=\gamma^2 \frac{k_*}{k_0} $ with $\alpha =150 $ (Dotted),  100 (Dot-Dashed),  50 (Dashed) and 9 (Solid). Requiring the dark matter mass is $m < H_\textrm{inf}/100 $ gives the Light Grey exclusion region. The Black region corresponds to the parameters for which the requirement for the vector dark matter to become non-relativistic prior to matter-radiation equality is violated, that is $ T_m<10^{-9} $ GeV, from Eq. (\ref{temp_mass_con}). The light red region indicates the parameters for which the energy density of the longitudinal component of the vector field generated during inflation is inconsistent with the constraints given in Eq. (\ref{long_con1}), while the black Dot-Dashed line indicates when the longitudinal mode explains $ 100\% $ of the currently observed dark matter energy density.}
	\label{allowed}
\end{figure}

Combining these relations we arrive at the following form for the predicted dark matter energy density today,
\begin{equation}
\rho_{\textrm{DM}}^{0}= \rho_{\textrm{DM}}^{\textrm{end}} \left( \frac{T_0}{T_m}  \right)^3 \left( \frac{ T_m}{T_\textrm{reh}} \right)^4~.
\label{DM_today1}
\end{equation}

Taking the observed value of the dark matter density today, given in Eq. (\ref{DM_today_obs}), we determine the required dark matter density at the end of inflation, from which we can place constraints on the model parameters. The vector dark matter energy density required at the end of inflation is given by,
\begin{equation}
\rho_{\textrm{DM}}^{\textrm{req}} =  9.6\cdot 10^{-48}~\textrm{GeV}^4  \left( \frac{T_m}{T_0}  \right)^3 \left( \frac{T_\textrm{reh}}{ T_m} \right)^4~,
\label{DM_end}
\end{equation}
which can be simplified as follows,
\begin{equation}
\rho_{\textrm{DM}}^{\textrm{req}} =  7\cdot 10^{57}~\textrm{GeV}^4  \left(\frac{10^{-10}~\textrm{GeV}}{ m}\right)\left(\frac{H_{\textrm{inf}}}{ 10^{12}~\textrm{GeV}}\right)^{5/2} ~,
\label{DM_end2}
\end{equation}
and must then be matched with Eq. (\ref{rhoVDM2}) to determine the allowed parameters. 

Through this matching, we determine the required $ \alpha $ parameter for a given dark matter mass and inflationary Hubble rate. This is depicted in Figure \ref{allowed}, in which the Light Red region represents a longitudinal mode energy density greater than $ 5\% $ of the observed dark matter energy density, and the Black Dot-Dashed line corresponds to it constituting $ 100\% $ of the dark matter. The Grey region is the upper bound on the vector mass from requiring $ m<H_\textrm{inf}/100 $, to ensure the mass term does not affect the resonance dynamics or play a role during inflation. The Black region represents the lower bound on the mass that ensures that the physical wavelength of the dark matter becomes smaller than the Compton wavelength before matter-radiation equality, see Eq. (\ref{temp_mass_con}). The Grey lines correspond to different values of the parameter $ \alpha=\gamma^2 \frac{k_*}{k_0} $~, namely, $\alpha =150 $ (Dotted),  100 (Dot-Dashed),  50 (Dashed) and 9 (Solid). Note that, if the dark matter mass is of the Stueckelberg type, the mass range may also be constrained by swampland constraints \cite{Reece:2018zvv}.

Through variation of the $ \alpha $ parameter, it is possible to have successful dark matter generation for a very large parameter range of dark matter masses and inflationary Hubble scales. To avoid significant longitudinal mode production, and to ensure $ T_m>10^{-9} $ GeV, we obtain an upper bound on the Hubble rate of $ H_\textrm{inf}<4.9\cdot 10^{12} $ GeV for $ m\simeq 2.6\cdot 10^{-12} $ GeV.

\begin{figure}[t]
	\centering
	\begin{subfigure}
		\centering
		\includegraphics[width=0.49\textwidth]{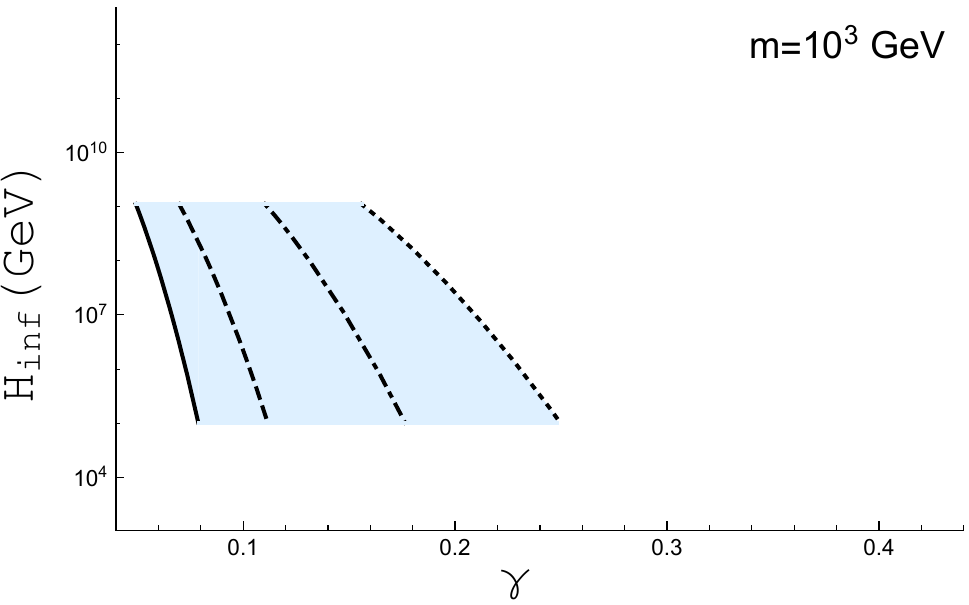}
	\end{subfigure}
	\hfill
	\begin{subfigure}
		\centering
		\includegraphics[width=0.49\textwidth]{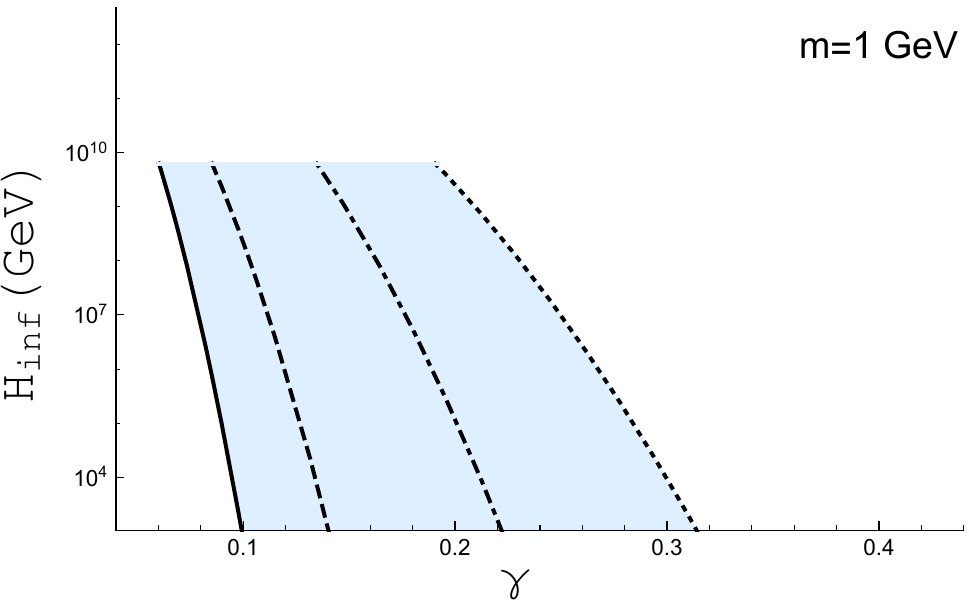}
	\end{subfigure}
	\begin{subfigure}
		\centering
		\includegraphics[width=0.49\textwidth]{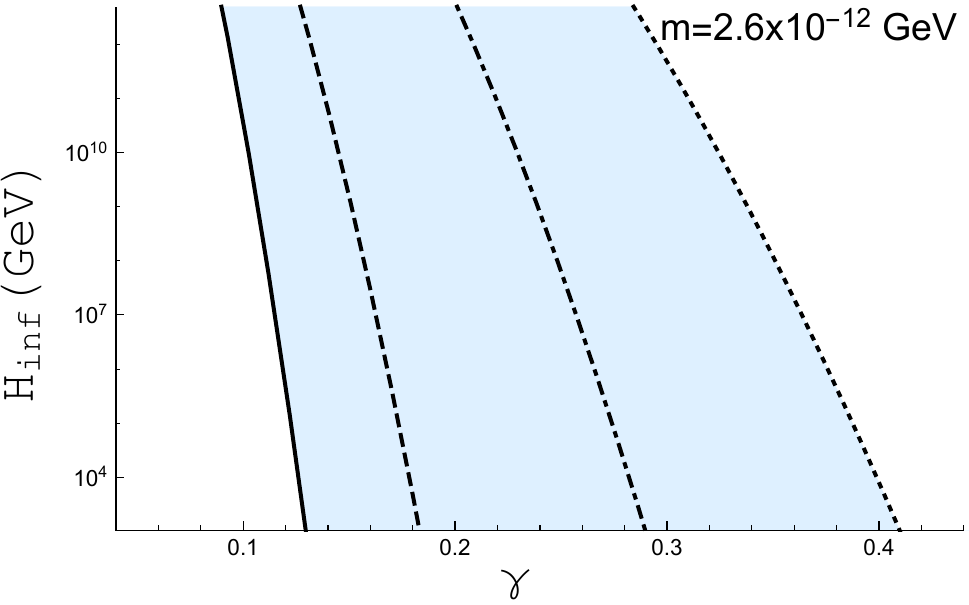}
	\end{subfigure}
	\hfill
	\begin{subfigure}
		\centering
		\includegraphics[width=0.49\textwidth]{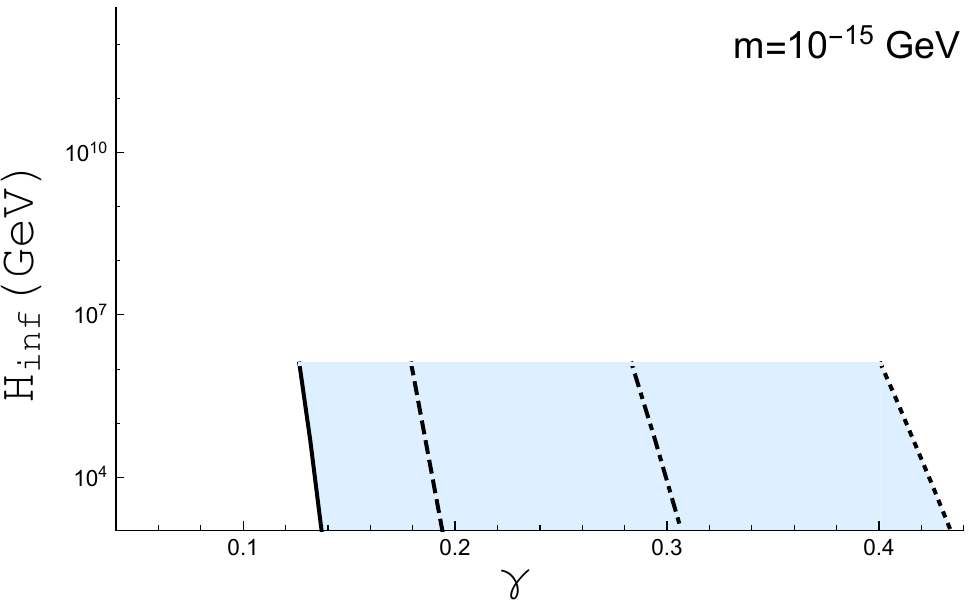}
	\end{subfigure}
	\caption{We show the $ \gamma $ and $ H $ parameter regions which lead to vector dark matter production consistent with the observed dark matter energy density today. The successful parameter regions for four different dark matter masses are depicted: (top left) $ m=10^{3} $ GeV, (top right) $ m=1 $ GeV, (bottom left) $ m=2.6\cdot 10^{-12} $ GeV, and (bottom right) $ m=10^{-15} $ GeV.  The black lines correspond to different values of $ \frac{k_*}{k_0} $~, namely $ \frac{k_*}{k_0}= 1000 $ (dotted), 2000 (dot-dashed), 5000 (dashed) and 10000 (solid). The horizontal cross sections denote lines of equal $ \alpha $.}
	\label{gamma_alpha}
\end{figure}

In this result, we have not taken into account the variation of the Hubble rate throughout inflation, and the possibility of a non-instantaneous reheating. Both of these would alter the required $ \alpha $ parameter for successful dark matter generation, and may shift the constraints on the $ m $ and $ H_\textrm{inf} $ parameter space.

In Figure \ref{gamma_alpha}, we depict the allowed regions of parameter space for $ \gamma $ and $ \frac{k_*}{k_0} $~, and the corresponding inflationary Hubble rate, for given dark matter masses. Each example given demonstrates the different constraints on the allowed parameter space of the Hubble rate as the dark matter mass varies, with $ m\simeq 2.6\cdot 10^{-12} $ GeV exhibiting the maximally allowed region. The $ m\simeq 10^{3} $ GeV  range is bounded from below by the maximal vector dark matter mass constraint, and above by significant longitudinal mode production.  The $ m\simeq 1 $ GeV  case is only constrained by longitudinal mode production, while $ m\simeq 10^{-15} $ GeV is only bounded from above by the requirement of the dark matter being non-relativistic before matter-radiation equality. The Black lines correspond to different values of the $\frac{k_*}{k_0} $ ratio, namely, 10000 (Solid), 5000 (Dashed), 2000 (Dot-Dashed), and 1000 (Dotted). Here we allow the IR cut-off $ k_0 $ to be a free parameter, but it may be associated with the mass of the vector dark matter. Similarly, the UV cut-off may be associated with the mechanism by which the dark matter mass is generated.


\subsection{Resonant Production During Inflation: $ k_*<H_\textrm{inf} $}

An important feature of this model is that it does not need to occur at the end of inflation. This means that the characteristic momenta scale of the vector dark matter at the end of inflation is not necessarily associated with the last modes to exit the horizon. Providing potential unique observable signatures related to the momentum distribution today and gravitational waves from the large energy density generation required, which contrasts other inflationary vector dark matter mechanisms.

We will now consider the case where $ k_* < H_{\textrm{inf}} $~, in which the resonance ends as the mode exits the horizon $ N_* $ e-folds before the end of inflation. After this occurs, the energy density will dilute  by $ e^{-4N_*} $ with the inflationary expansion. Additionally, the physical momenta of the vector dark matter at the end of inflation will be smaller than $ H_{\textrm{inf}} $ by a factor of $ e^{-N_*} = k_*/H_{\textrm{inf}}$~. This means that the temperature $ T_m $ will be increased by a factor of $ e^{N_*}$, such that the vector dark matter becomes non-relativistic earlier, enhancing the dark matter density that would be present today. Thus, the required dark matter energy density at the end of inflation is reduced, and is given by,
\begin{equation}
	\rho_{\textrm{DM}}^{\textrm{req}} =  7\cdot 10^{57}~\textrm{GeV}^4  \left(\frac{10^{-10}~\textrm{GeV}}{ m}\right)\left(\frac{H_{\textrm{inf}}}{ 10^{12}~\textrm{GeV}}\right)^{5/2} \frac{k_*}{H_{\textrm{inf}}} ~.
	\label{DM_end_*}
\end{equation}

In considering this case, we must be careful that the energy density of the dark matter generated at $ N_* $ e-folds before the end of inflation is not too large such that it disrupts the background evolution. By taking the requirement that $ \rho_* \ll 3M_p^2 H_\textrm{inf} $~, we determine the earliest point in inflation that the vector dark matter resonance safely occurs. This lower bound on the earliest the resonance is allowed to occur, while simultaneously explaining the observed dark matter and not disrupting inflation, is given by the following,
\begin{equation}
 \frac{k_*}{H_{\textrm{inf}}}> 9.3 \cdot 10^{-7}\frac{H_{\textrm{inf}}^{1/6}}{ m^{1/3}} ~,
	\label{DM_limit}
\end{equation}
where we have assumed that the energy density of the vector dark matter constitutes less than $ 5\% $ of the total energy density of the universe at horizon exit, $ N_* $ e-folds before the end of inflation.

Using this relation, we are able to determine the earliest possible point in inflation that the resonance for the $ k_* $ mode can occur. Larger dark matter masses allow for earlier resonances, so that the maximal number of e-folds before the end of inflation will result from the maximum $ m $ that ensures the mass term plays no role during inflation, $m = k_*/100 $. This gives the following relation,
\begin{equation}
\frac{k_*}{H_{\textrm{inf}}}>6.6 \cdot 10^{-6}~\left(\frac{1.9\cdot 10^9 ~ \textrm{GeV}}{ H_{\textrm{inf}}} \right)^{1/8}~,
	\label{DM_limitmax}
\end{equation}
where for Hubble rates greater than $ H_{\textrm{inf}}\simeq 1.9\cdot 10^9 $ GeV, the longitudinal mode energy density becomes greater than $ 5\% $. This means the resonance can occur as early as approximately 11.9 e-folds before the end of inflation for $ H_{\textrm{inf}}\simeq 1.9\cdot 10^9 $ GeV and $ m\simeq 125 $ GeV, and 10.1 e-folds for $ H_{\textrm{inf}}\simeq 10^3 $ GeV and $ m\simeq 4\cdot 10^{-4} $ GeV. This maximal case allows for an approximate $ 10^4-10^5 $ times larger characteristic length scale for the dark matter today compared to the $ k_*=H_\textrm{inf} $ case.

Consideration of the constraint from longitudinal mode production, in Eq. (\ref{long_con}), is necessary for inflationary Hubble rates greater than $ H_{\textrm{inf}}\simeq 1.9\cdot 10^9 $ GeV. Incorporating this constraint into Eq. (\ref{DM_limit}) gives the following lower bound on $ \frac{k_*}{H_{\textrm{inf}}} $~,
\begin{equation}
\frac{k_*}{H_{\textrm{inf}}}> 6.6 \cdot 10^{-6} \left(\frac{H_{\textrm{inf}}}{1.9\cdot 10^9  ~\textrm{GeV}} \right)^{3/2}~,
\label{DM_limit_long}
\end{equation}
where we have assumed that the longitudinal component makes up no more than $ 5\% $ of the dark matter energy density.

The lower physical momentum at the end of inflation means that for a given vector dark matter mass, it will become non-relativistic earlier than in the $ k_*=H_\textrm{inf} $ case. This may allow for the possibility of lower mass vector dark matter candidates than in the previous scenario. We now consider the limit imposed on the dark vector mass when requiring $ T_m>10^{-9} $ GeV, given by,
\begin{equation}
m>2.6 \cdot 10^{-12} ~\textrm{GeV}~  \left(\frac{H_\textrm{inf}}{4.9 \cdot 10^{12}~\textrm{GeV}}\right)^{1/2} \frac{k_*}{H_{\textrm{inf}}}~,
\label{temp_mass_con1}
\end{equation}
which upon combining with the lower bound on $ \frac{k_*}{H_{\textrm{inf}}} $ in Eq. (\ref{DM_limit}),
\begin{equation}
m>2.4\cdot 10^{-12} ~\textrm{GeV}~  \sqrt{\frac{H_\textrm{inf}}{4.9 \cdot 10^{12}~\textrm{GeV}}} ~,
\end{equation}
where we find that due to energy density constraint during inflation, there is a negligible expansion in the allowed range of vector dark matter masses compared to the $ k_*=H_\textrm{inf} $ scenario. The enhancement of the dark matter energy density from the lowered characteristic momenta is not sufficient to offset the additional dilution induced after horizon crossing early in the inflation epoch. Thus, the region of allowed parameter space is similar to the $ k_*=H_\textrm{inf} $ case, with the exception of the upper bound we impose for analytic consistency, $ m<k_*/100 $. The required $ \alpha $ parameter for different $ m $ and $ H_\textrm{inf} $ parameter sets is now also dependent on the variation of $ k_*/H_\textrm{inf}$~.


\section{Phenomenological Implications}
\label{phenom}

Now that we have seen how this mechanism leads to the successful production of the observed dark matter energy density, we will now describe the phenomenological implications of this model. Importantly, the required dynamics of the resonant production during inflation does not preclude the possibility of small couplings between the vector dark matter and the Standard Model (SM) sector. A particularly important example is kinetic mixing between the vector dark matter and the photon. If such a coupling exists, then it would provide prospects for the detection of the vector dark matter at terrestrial experiments or through observational constraints. The extended range of allowed masses for the vector dark matter means that this mechanism will be probed at a wide array of current experimental searches for dark photons \cite{Fabbrichesi:2020wbt,Caputo:2021eaa}.

The dark sector may include more than just the vector dark matter candidate. In particular, the mass generation mechanism for the dark matter could be induced by a non-zero vacuum expectation value of a dark scalar field. A Higgs portal interaction between this scalar and the Higgs would then provide interactions with the SM sector that are testable at terrestrial experiments. Additionally, gravitational wave signals produced by the dark phase transition associated with this scalar taking its non-zero vacuum expectation value may be observable at future gravitational wave detectors \cite{Kosowsky:1991ua,Kamionkowski:1993fg,Huber:2008hg}.

The dark matter energy density produced in this scenario is carried in the transverse components of the vector field, and is made up of both circular polarisations. This non-chiral nature is in contrast to the tachyonic production mechanism \cite{Bastero-Gil:2018uel} which produces only a single polarisation, and to the case of longitudinal production \cite{Graham:2015rva}. It may be possible to probe this feature at future experiments, providing an additional method to differentiate these mechanisms \cite{Wolf:2018xlz,Alonso:2018dxy}.

It is also important to note that the detection of this vector dark matter candidate will also illuminate aspects of the inflationary scenario. Depending on the form of the coupling between the inflaton and the vector dark matter, the characteristic length scale and mass of the dark matter will provide insights into the form of the inflationary potential. Providing complementary information to that found by improved CMB searches and measurements of the tensor-to-scalar ratio. There may also be observable non-gaussianities associated with the form of the potential that induces the required oscillatory behaviour \cite{Chen:2008wn,Flauger:2010ja,Chen:2010bka,Leblond:2010yq}.

Two key ways in which this scenario provides unique phenomenological implications, besides those mentioned above, are the energy density spectrum and gravitational waves. Firstly, the resonant amplification produces a sharply peaked spectrum around the characteristic mode $ k_* $~. Today, the two regimes considered in the previous sections lead to dark matter energy densities with different characteristic length scale relationships with the inflationary Hubble rate. Additionally, the range of possible length scales is increased considerably in the $ k_*<H_\textrm{inf} $ due to the much earlier production process. Quantitatively, in the case of $ k_*=H_\textrm{inf} $~, the characteristic length scales today are,
\begin{equation}
\mathcal{O}(10) \textrm{~cm~} <\frac{1}{k_*}<  \mathcal{O}(10) \textrm{~km}~,
\end{equation} 
similarly to the case in \cite{Bastero-Gil:2018uel}. However, when we allow for $ k_*=H_\textrm{inf} $~, the upper bound is significantly relaxed, giving,
\begin{equation}
\mathcal{O}(10) \textrm{~cm~} <\frac{1}{k_*}<  \mathcal{O}(10^6) \textrm{~km}~,
\end{equation} 
where the upper bound is derived from the constraint on the maximal number of e-folds before the end of inflation that the resonance production can occur. This is due to the additional redshift induced by the earlier production of the vector dark matter energy density during inflation. This range of values is much larger than that which is possible in Ref. \cite{Bastero-Gil:2018uel}, but smaller than that characteristic of the longitudinal case \cite{Graham:2015rva}. Other contrasting observables in combination with these results will help to differentiate these inflationary vector dark matter scenarios.

There are three main sources of gravitational wave signatures in this model - the inflationary tensor-to-scalar ratio, those associated with the possible mass generation mechanism for the vector dark matter, and gravitational waves sourced from the gauge field production. Firstly, the predicted tensor-to-scalar ratio is related to the inflationary Hubble rate through the slow roll relations, allowing the maximal value to be determined from our upper bound on $ H_\textrm{inf} $~. In the scenario where we have a Stueckelberg mass term, the upper limit on the inflationary Hubble rate leads to the following upper bound on the tensor-to-scalar ratio,
\begin{equation}
r\lesssim 4\cdot 10^{-4}
\end{equation}
for  $m\simeq 2.6 \cdot 10^{-12}$ GeV and $H_{\textrm{inf}}=4.9\cdot10^{12}$ GeV. This is below the projected sensitivity of the LiteBIRD telescope \cite{Hazumi:2019lys}. Thus, this upcoming experiment will only constrain the parameter region in which the vector dark matter does not have a Stueckelberg mass term during inflation. 

If the mass is instead generated by a dark scalar field taking a vacuum expectation value later in the evolution of the universe, then the associated phase transition could generate observable gravitational waves. This will depend on the various model parameters and corresponding mass scale. 

The enhanced production of gauge fields in inflationary models is known to be able to generate unique gravitational signatures \cite{Cook:2011hg,Ben-Dayan:2016iks,Barrie:2020kpt, Barrie:2021orn}, which can provide an additional avenue for observational testing. The rapid growth of the vector dark matter energy density during the resonance period in our scenario will lead to the production of gravitational waves. These will be non-helical and exhibit a power spectrum strongly associated with the peak resonance mode $ k_* $~, leading to a differentiating feature compared to other vector dark matter scenarios. Additionally, observation of the gravitational waves will act as a probe of both the dark matter and the inflationary potential. 

For the $ k_*=H_\textrm{inf} $ case, the percentage of the total universe energy density during the resonance enhancement is significantly smaller than is necessary in the case when $ k_*<H_\textrm{inf} $~. Thus, the $ k_*<H_\textrm{inf} $ scenario offers the best opportunity for generating an observable gravitational wave signal that exceeds the inflationary vacuum contributions. The resultant gravitational waves will be non-chiral and associated with the characteristic mode $ k_* $~, allowing for matching with the characteristic length scale of the dark matter today. Thus, these gravitational wave signals, in combination with the other phenomenological implications discussed above, demonstrate the wide range of tests through which this mechanism will be probed in the near future. Successful detection of these predicted signals will provide opportunities to illuminate the connections between the nature of dark matter and the properties of the inflationary mechanism.


\section{Conclusion}
\label{conc}

We have demonstrated a new mechanism for generating the observed dark matter energy density through the resonant production of a vector field during inflation. The resonance is induced by a coupling between the inflaton and the vector dark matter of the form $ f(\varphi)X_{\mu\nu}\tilde{X}^{\mu\nu} $~. Both chiralities of the vector field are generated, alongside a minimal longitudinal component. A wide range of dark matter masses and inflationary Hubble rates lead to the successful prediction of the observed dark matter density. Depending on the coupling structure, the resonant production can occur as early as $ \sim 11.9$ e-folds before the end of inflation, allowing for a different range of characteristic length scales of the dark matter today compared to other inflationary dark matter generation models. The early production case may also generate unique gravitational wave signatures due to the larger vector field energy densities required to compensate for the subsequent inflationary dilution. This will require a detailed numerical analysis to determine the expected signals.

We have assumed that the longitudinal component constitutes a negligible contribution to the total dark matter density, but this is not necessary. It may be that the transverse and longitudinal components both contribute significantly, which would weaken the upper bound we have applied on the maximal inflationary Hubble rate. Additionally, we have assumed an instantaneous reheating and standard cosmological evolution, which if false would alter the longitudinal bound as well as the required $ \alpha $ parameter for given vector dark matter mass and inflationary Hubble rate parameters.

However, the longitudinal mode constraint applies only if the vector dark matter has a Stueckelberg mass term. If it is not, then the allowed parameter space can be extended to larger inflationary Hubble rates, up to the current upper bound from CMB observations. The corresponding upper bound on the $ m $ and $ H_\textrm{inf} $ parameter space would instead come from the lower bound on the parameter $ \alpha>9 $ for the consistency of our analytical results.

Although we have used the above form of coupling, it is not necessary that this be the inflaton to vector dark matter coupling and that an exact Mathieu resonance appears. This example has been used to demonstrate the ease with which this mechanism generates the observed dark matter energy density, and thus, less efficient oscillatory coupling scenarios should also be able to succeed. An additional advantage of this coupling and the associated resonance behaviour is that a large coupling is not required for significant vector dark matter production.

The vector dark matter candidate could exhibit couplings to the SM particles through kinetic mixing with the photon. This possibility allows for many avenues to probe the properties of the vector field terrestrially, in concert with complementary observational tests. Additionally, this model can be embedded in a dark Higgs mechanism scenario, providing extra routes for probing the dark sector such as Higgs portal couplings and gravitational waves from an associated phase transition. Combining each of these detection possibilities we arrive at a  unique combination of observational signatures that will allow not only the elucidation of the nature of dark matter but also of the inflationary dynamics.


\section*{Acknowledgements}

This work was supported  by IBS under the project code, IBS-R018-D1. The author would like to thank Shigeki Matsumoto and Kavli IPMU for their support during the completion of this work.

\end{document}